\documentclass{moriond}

\def\Journal#1#2#3#4{{#1} {\bf #2}, #3 (#4)}

\def\be{\begin{equation}}
\def\ee{\end{equation}}
\def\bea{\begin{eqnarray}}
\def\eea{\end{eqnarray}}

\def\CQG{\it Class. Quant. Grav.}
\def\etal{\textit{et al.}~}

\newcommand{\Photo}{\includegraphics[height=35mm]{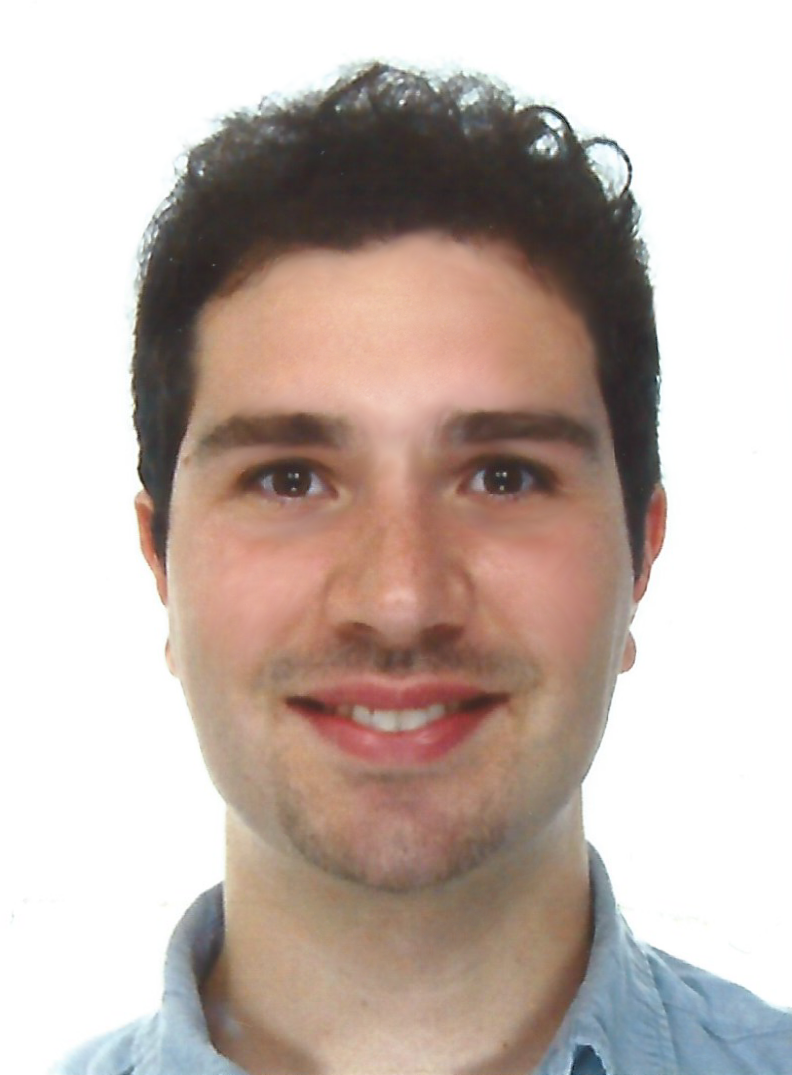}}

\begin{document}
\vspace*{4cm}
\title{\texttt{BRiSTOL} - a Band-limited RMS Stationarity Test Tool for Gravitational Wave Data}

\author{ F. DI RENZO,$^{(*)}$ F. FIDECARO, M.RAZZANO and N. SORRENTINO}

\address{
	Universit\`a di Pisa, Dipartimento di Fisica, I-56127 Pisa, Italy\\
	Istituto Nazionale di Fisica Nucleare, Sezione di Pisa, I-56127 Pisa, Italy\\[2ex]
	$^{(*)}\mathrm{Corresponding\ author}$ francesco.direnzo@df.unipi.it
	}

\maketitle\abstracts{
	Common techniques in Gravitational Wave data analysis assume, to some extent, the stationarity and Gaussianity of the detector noise.
	These assumptions are not always satisfied because of the presence of short duration transients, namely glitches, and other slower variations in the statistical properties of the noise, which might be related to malfunctioning subsystems.
	We present here a new technique to test the stationarity hypothesis with minimal assumptions on the data, exploiting the band-limited root mean square and the two-samples Kolmogorov-Smirnov test.
	The outcome is a time-frequency map showing where the hypothesis is to be rejected.
	This technique was used as part of the event validation procedure for assessing the quality of the LIGO and Virgo data during O3.
	We also report on the applications of the test to both simulated and real data, highlighting its sensitivity to various kinds of non-stationarities.
	}

\section{Introduction}

Gravitational Waves (GWs) are propagating ``ripples'' in the fabric of space-time, produced by the coalescence of compact binary star systems (CBCs) or other violent phenomena in the Universe~\cite{Sathyaprakash2009}. Interferometric GW detectors, like Advanced LIGO~\cite{TheLIGOScientific:2014jea} and Advanced Virgo~\cite{TheVirgo:2014hva}, measure their transit from the differential strain induced on the detector arms. Many sources of \emph{noise}, either of environmental or instrumental origin, can produce a similar effect. This noise is conveniently described as a \textit{stochastic} (or \textit{random}) \textit{process}, and our ability to extract the information on the astrophysical signal is determined by how good we know the statistical properties of the noise~\cite{LIGOScientific:2019hgc}.

Common GW data analysis techniques assume that the noise process can be modeled as a  \textit{stationary} and \textit{Gaussian} one~\cite{Jaranowski:2005hz}. Stationarity implies that its statistical properties do not vary over time, and can then be estimated from a single, sufficiently long, realization of the process. Gaussianity ensures that the second order distribution moments are sufficient to fully characterize it. In particular, the power spectral density (PSD) of the Gaussian process is the sole quantity needed to fully represent its statistical properties.

When these assumptions cease to be valid, the noise characterization gets complicated, the techniques used within the stationary and Gaussian assumption are no more optimal~\cite{Zackay:2019kkv}, and the corresponding estimates of the GW source properties are altered~\cite{Edy:2021par}. Moreover, noise non-stationarities are often the manifestation of some subsystem malfunction in the detector~\cite{Aasi:2012wd}. For these reasons, it is important to identify them, both for the assessment of the data quality in correspondence of a candidate event and from the point of view of the characterization and improvement of the detector.


\section{A Stationarity Test Based on the Signal Band-Limited RMS}

We present here a new test of stationarity based on the signal Band-Limited RMS (BLRMS) that we have called \texttt{BRiSTOL}: Band-RMS Stationarity Test toOL~\cite{france}. 

\subsection{Estimation of the Signal BLRMS}
The BLRMS is the average power of a signal in a limited frequency band $[f^{\min} ,f^{\max}]$:
\begin{equation}\label{eq:BLRMS}
\mathit{BLRMS}\big(t;[f^{\min} ,f^{\max}]\big)=\sqrt{\frac{1}{f^{\max}-f^{\min}}\int_{f^{\min}} ^{f^{\max}} \hat{S}(f;t) df}
\end{equation}
where $\hat{S}(f;t)$ is an estimate of the signal PSD referred to the time $t$.

Using this quantity provides two advantages with respect to standard analyses based on the PSD variability. 
Firstly, the highly correlated GW detector noise has regions of its spectrum characterized by very diverse behaviors. There are parts of it with characteristic features, like \emph{spectral lines}, bumps and other structures where one would need a better frequency resolution, and regions where the spectrum is \emph{flatter}, described for example by some characteristic power law. Averaging over the latter frequency regions has a variance reduction effect similar to that obtained with the Welch method~\cite{welch1967} in the time domain.

Secondly, many noise sources affect specific frequency bands only. If one knows a band division of the spectrum able to bound them, these bands can be used for the BLRMS estimation. Then, once a non-stationary noise component stands out in a specific band, this can be immediately associated to a known possible source.

\begin{figure}
\hspace{-3mm}
\includegraphics[height=2.5in]{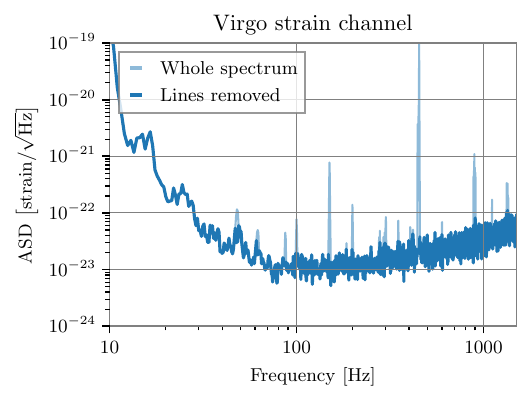}
\includegraphics[height=2.5in]{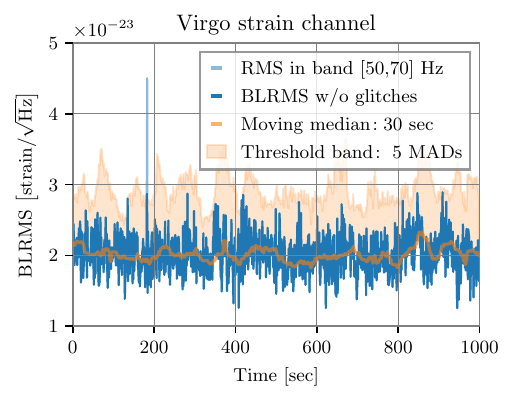}
\caption{\label{fig:removals}Left: effect of spectral line removal on the Amplitude Spectral Density (ASD) of Virgo. Right: glitch removal from a Virgo BLRMS time series, where a glitch at about $200$ seconds from the start is removed by the thresholding MAD (orange band). These plots have been created with O3a open data of April 5, 2019~\protect\cite{gwosc}.}
\end{figure}

\subsection{Modified BLRMS Algorithm}

The algorithm to compute the BLRMS from Eq.~\ref{eq:BLRMS} can be improved with two additions to make it more effective for the purpose of GW detector noise studies. Firstly, if a spectral line is contained in the band where we are estimating the BLRMS, this is going to dominate the result. This is sometimes undesirable, especially if one is mostly interested in the non-stationarities of the \textit{noise floor}. 
For this reason, we have included in the method for computing the BLRMS the possibility to automatically identify spectral lines and remove them with a technique similar to the one described in~\cite{Acernese:2005fi}.

Secondly, GW data often presents \emph{glitches}. These are rapid excess of noise, with typical duration shorter than one second. This feature makes them similar to \emph{burst} GW signals. The best algorithms to identify them are the so called \emph{Event Trigger Generators} (ETGs)~\cite{Robinet:2020lbf}. Then, to focus on slower non-stationarities, usually not targeted by these ETGs, we may want to exclude the glitches from our stationarity test. For this reason, we added the possibility to identify and remove them with an algorithm that computes the moving median of the BLRMS' time series and applies a threshold based on their Median Absolute Deviation (MAD).

Examples of spectral line and glitch removal on Virgo O3a data are shown in Fig.~\ref{fig:removals}.

\subsection{BLRMS-based Stationarity Test}

We have based our stationarity test on the modified BLRMS computation described above. Firstly, for every frequency band, we divide the BLRMS time series in consecutive chunks containing a statistically significant number of points. Then, we make use of specific \emph{metrics} to test the stationarity hypothesis comparing the BLRMS estimates in consecutive chunks. 

\begin{figure}
	\centerline{\includegraphics[width=6.5in]{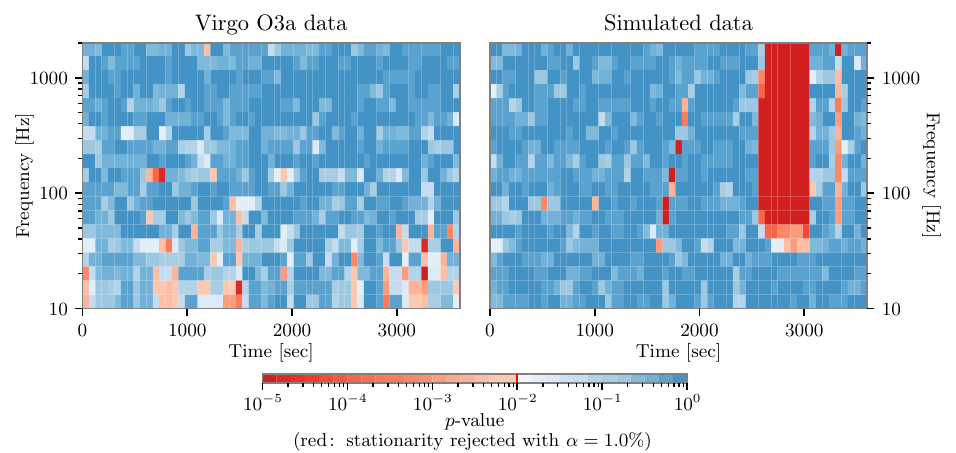}}
	\caption{\label{fig:BLRMStest}Time-frequency map of the $p$-values for the null hypothesis of stationarity obtained with \texttt{BRiSTOL}. Red regions are times and frequency bands where this hypothesis should be rejected at a significance level $\alpha=1\%$. Left: application to real Virgo O3a open data of April 5, 2019. Right: simulated colored Gaussian Virgo data, randomly generated from the ASD in Fig.~\ref{fig:removals} with the addition of some noise non-stationarities.} 
\end{figure}

A particularly balanced metric, useful for having an algorithm sensitive to a large variety of non-stationarities, is the one provided by the \emph{two-samples Kolmogorov-Smirnov test}~\cite{kolmogoroff1941confidence}, based on the maximum absolute difference between the BLRMS empirical c.d.f.s in neighboring chunks, respectively with $n$ and $n'$ data points, $\{x_i\}_{i=1,...,n^{(\prime)}}$:
\begin{equation}
\hat{F}_{n}(x)=\frac{1}{n}\sum_{i=1}^{n}I(x_i\leq x)\label{eq:Empirical_cdf}
\end{equation}
where $I(\ldots )$ is the \emph{indicator} or \emph{characteristic function}, equals to $1$ if its argument is true, zero otherwise. The previous quantity counts the proportion of the sample points below level $x$. The test statistic is therefore:
\begin{equation}\label{eq:KS_test_statistic}
\mathit{KS}_{nn'}=\sqrt{\frac{nn'}{n+n'}}\sup_{x}\left|\hat{F}_{n}(x)-\hat{F}_{n'}(x)\right|.
\end{equation}
With this normalization, the test statistic asymptotically approaches a \emph{Kolmogorov distribution}, which is independent on the distribution of the data (\emph{Kolmogorov Theorem}). Also, $p$-values, the probabilities of obtaining more extreme values from the same distribution, are available for this statistic and can be reported in a time-frequency map where to verify the stationarity hypothesis. An example of this representation for the test outcome is shown in Fig.~\ref{fig:BLRMStest}.

\section{Applications and Conclusions}
In Fig.~\ref{fig:BLRMStest} we reported two examples of application of \texttt{BRiSTOL}. The time-frequency maps show the $p$-values for the stationarity hypothesis in various time chunks and frequency bands. Red regions are where stationarity should be rejected at a significance level $\alpha=1\%$, that is, the $p$-value there is smaller than this threshold. On the left, we reported the application to real O3a Virgo data, while on the right we generated a (stationary) colored Gaussian noise from the ASD in Fig.~\ref{fig:removals}, to which we added some noise features: a sinusoid at frequency $90$ Hz from time $500$ to $1000$ sec,\footnote{Notice that a sinusoidal signal is stationary. The only features highlighted by the colormap in Fig.~\ref{fig:BLRMStest} are indeed its onset and offset, respectively at time $500$ and $1000$ seconds from the start.} a moving line with exponentially varying frequencies between $10$ and $1000$ Hz from $1500$ to $2000$ sec, a ``ramp'' of white noise with increasing amplitude from $2500$ and $3000$ sec, and lastly a $5$ seconds ``burst'' (covering all frequency bands) at time $3250$ sec. All these features have amplitude (the maximum one for the ramp and the burst) equals to $10^{-22}$ in strain units.

\texttt{BRiSTOL} has been used for the validation of candidate events during O3 and for detector characterization studies. Thanks to the  minimal assumptions on the data, namely that the possible non-stationarities can show up in the second order moments,\footnote{A subsequent Gaussianity test can also reveal if this is sufficient to exclude non-stationarities at higher order.} this test can be used as a first step for assessing data quality. If non-stationary regions are identified, further analyses are triggered to investigate the possible origin of them or, in the case of candidate event validation, to check whether they can be detrimental for the estimations on the GW source properties.

\end{document}